\documentclass[aip,pof,reprint,unsortedaddress,amsmath,amssymb]{revtex4-1}

\usepackage{graphics,graphicx}

\begin{document}

\title{Statistical mechanics and large-scale velocity fluctuations of turbulence}

\author{Hideaki Mouri}
\email{hmouri@mri-jma.go.jp}

\author{Akihiro Hori}
\altaffiliation[Also at ]{Meteorological and Environmental Sensing Technology, Inc.}

\author{Yoshihide Kawashima}
\altaffiliation[Also at ]{Meteorological and Environmental Sensing Technology, Inc.}

\author{Kosuke Hashimoto}
\altaffiliation[Also at ]{Meteorological and Environmental Sensing Technology, Inc.}

\affiliation{Meteorological Research Institute, Nagamine, Tsukuba 305-0052, Japan}

\date{18 November 2011}


\begin{abstract}
Turbulence exhibits significant velocity fluctuations even if the scale is much larger than the scale of the energy supply. Since any spatial correlation is negligible, these large-scale fluctuations have many degrees of freedom and are thereby analogous to thermal fluctuations studied in the statistical mechanics. By using this analogy, we describe the large-scale fluctuations of turbulence in a formalism that has the same mathematical structure as used for canonical ensembles in the statistical mechanics. The formalism yields a universal law for the energy distribution of the fluctuations, which is confirmed with experiments of a variety of turbulent flows. Thus, through the large-scale fluctuations, turbulence is related to the statistical mechanics.
\end{abstract}

\maketitle


\section{Introduction} \label{S1}

\begin{figure}[b]
\rotatebox{90}{
\resizebox{4.1cm}{!}{\includegraphics*[3.6cm,2.8cm][16.3cm,27cm]{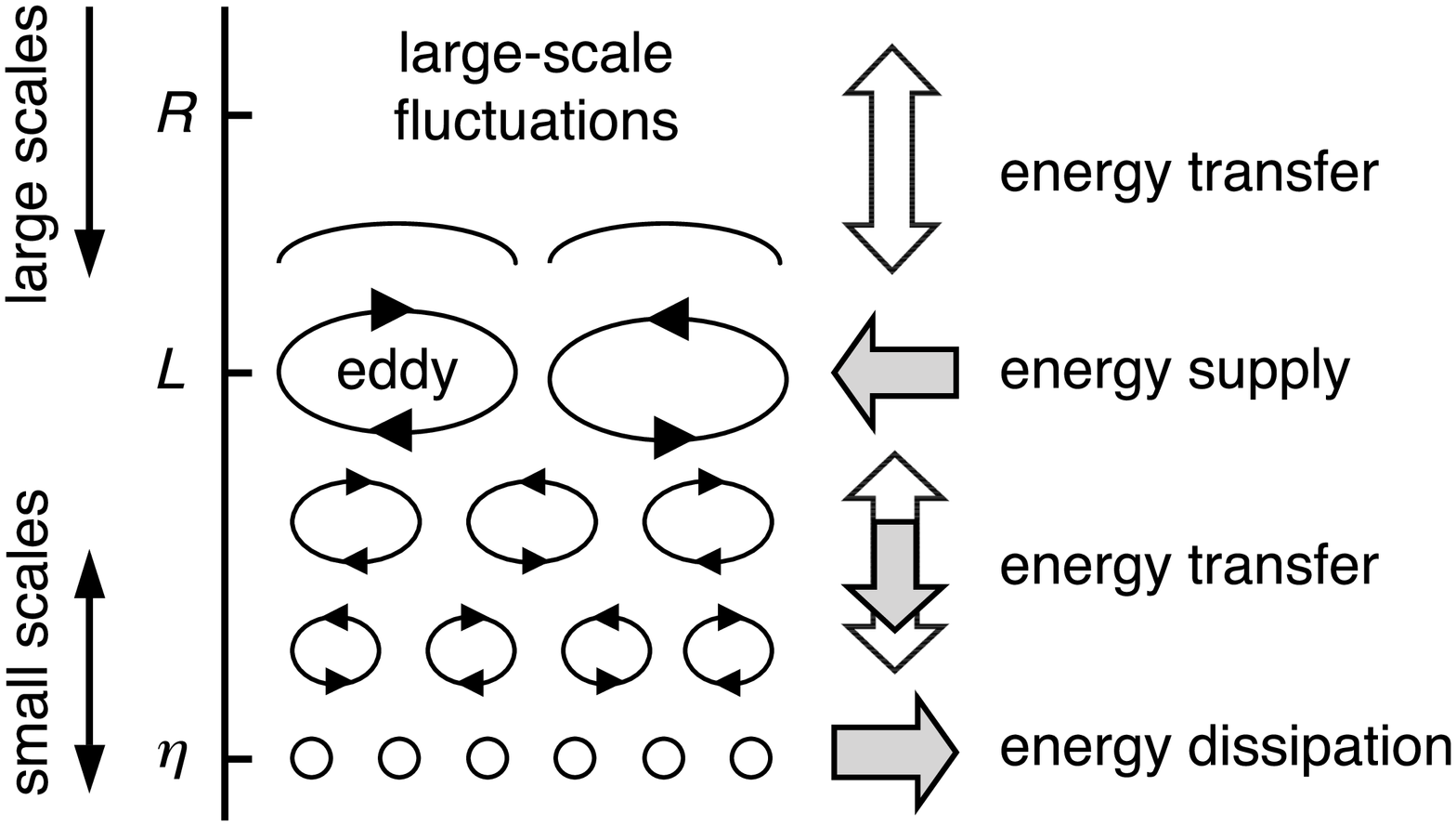}}
}
\caption{\label{f1} Sketch of three-dimensional stationary turbulence. The open arrows indicate the local and instantaneous energy transfer. The filled arrows indicate the net energy transfer along with the energy supply and the energy dissipation.}
\end{figure}

Turbulence is induced by supplying kinetic energy at some scale $L$. This energy could be transferred to both the larger and the smaller scales.\cite{dr90,ok92} However, as sketched in Fig.~\ref{f1}, the energy is on average transferred to smaller and smaller scales because it is eventually dissipated into heat at the smallest scale, i.e., the Kolmogorov length $\eta$. The energy transfer from $L$ to $\eta$ consists of many random steps, each of which occurs preferentially between neighboring scales.\cite{dr90,ok92,f95} Hence, although motions at the scale $L$ depend on the flow configuration for the energy supply, such dependence is lost during the energy transfer. The resultant small-scale motions exhibit universal features, which have been studied as a representative of spatially correlated nonequilibrium fluctuations.\cite{f95}

The kinetic energy could be transferred to scales much larger than $L$ and could cause velocity fluctuations there (Fig.~\ref{f1}). Also in these large-scale fluctuations, we expect some universality. To lose dependence on the flow configuration, the energy transfer could have a sufficient number of random steps. Any step prefers to occur between neighboring scales because such scales alone interact in a coherent manner.\cite{f95}

If the turbulence is stationary and is homogeneous at least along one direction, we expect that the large-scale fluctuations are analogous to thermal fluctuations in an equilibrium state described in the standard textbooks\cite{k65,ll80,c85} of the statistical mechanics. The stationarity implies that no net energy is transferred across the large scales, while the homogeneity implies that no net energy is transferred in space along that direction. This is analogous to the case of the thermal equilibrium at constant temperature, where occurs no net heat transfer.\cite{k65,ll80,c85} In addition, at the large scales of the homogeneous turbulence, we could ignore any of the spatial correlations. Then, the fluctuation energy is additive. Its value for a large-scale region is the sum of its values for the yet large parts of the region that are not correlated at all.\cite{k65,ll80} The large-scale fluctuations are thus considered as a collection of many distinct motions. This is again analogous to the case of the thermal fluctuations, which have many degrees of freedom.

The large-scale fluctuations of turbulence are known to be significant, regardless of the flow configuration,\cite{po97,c03,m06,mht09} as considered by Landau.\cite{ll59,m06,f95,k74} However, their details are still not known. Experimentally or numerically, any detailed study needs long data for many realizations of the large scales. Such data have not been available. The situation is nevertheless improving,\cite{mht09} owing to improvements in experimental technologies.

By assuming the universality and the additivity, we describe the equilibrium large-scale fluctuations of the stationary homogeneous turbulence in a thermostatistical formalism, i.e., a formalism that has the same mathematical structure as used for the statistical mechanics.\cite{o49,s72,bo08} The formalism is confirmed with long experimental data of a variety of turbulent flows obtained in a wind tunnel. We thereby demonstrate that turbulence is related through its large-scale fluctuations to the statistical mechanics.

\section{Configuration and Coarse Graining} \label{S2}

Let us consider a lateral velocity $v(x)$ along some one-dimensional cut $x$ of three-dimensional stationary turbulence. The longitudinal velocity $u(x)$ is to be also considered, by replacing $v$ with $u$ in the following descriptions. The turbulence is assumed to be homogeneous along the one-dimensional cut. The average $\langle v \rangle$ is subtracted so as to have $\langle v \rangle = 0$ anywhere below. As a characteristic scale $L$ of the energy supply, we use the correlation length of the local energy $v^2$. The usual definition is
\begin{subequations}
\label{eq1}
\begin{equation}
\label{eq1a}
\tilde{L}_{v^2} = \frac{\int^{\infty}_0 \langle [v^2(x+r) - \langle v^2 \rangle][v^2(x) - \langle v^2 \rangle] \rangle \, dr}
            {\langle (v^2  - \langle v^2 \rangle)^2 \rangle} ,
\end{equation}
but our definition for later convenience is
\begin{equation}
\label{eq1b}
L_{v^2} = \frac{\langle (v^2  - \langle v^2 \rangle)^2 \rangle}{2\langle v^2 \rangle^2} \tilde{L}_{v^2}. 
\end{equation}
\end{subequations}
We have $L_{v^2} = \tilde{L}_{v^2}$ in a special case where the distribution of $v$ is Gaussian, $\langle v^4 \rangle = 3 \langle v^2 \rangle^2$.

The one-dimensional cut is divided into segments with length $R$. For each segment, the center of which is tentatively defined as $x_{\rm c}$, the energy $v^2$ is averaged as
\begin{equation}
\label{eq2}
v_R^2(x_{\rm c})=\frac{1}{R} \int^{+R/2}_{-R/2} v^2(x_{\rm c}+x) \, dx .
\end{equation}
We focus on this coarse-grained energy. The mean square of $v_R^2$ around its average $\langle v_R^2 \rangle = \langle v^2 \rangle$ is\cite{k65,r54}
\begin{subequations}
\label{eq3}
\begin{align}
\label{eq3a}
&
\langle (v_R^2 - \langle v_R^2 \rangle )^2 \rangle \\
&
= \frac{2}{R^2} \int^R_0  
                          (R-r)
                          \langle [v^2(x+r) - \langle v^2 \rangle][v^2(x) - \langle v^2 \rangle] \rangle dr , \nonumber
\end{align}
where the same symbol $\langle \cdot \rangle$ is used to denote averages over the positions $x$ and over the segments. Since the energy is supplied at around the scale $L_{v^2}$, we assume that any $n$-point spatial correlation of $v^2$ decays fast enough to become negligible at $r \gg L_{v^2}$. In other words, we assume additivity of $R v_R^2$ at $R \gg L_{v^2}$. Then, Eqs.~(\ref{eq1}) and (\ref{eq3a}) yield
\begin{equation}
\label{eq3b}
\langle (v_R^2 - \langle v_R^2 \rangle )^2 \rangle
=
\frac{4L_{v^2}}{R} \langle v^2 \rangle^2 
\quad
\mbox{at}
\quad
R \gg L_{v^2} .
\end{equation}
\end{subequations}
The assumption also implies $\langle (v_R^2 - \langle v_R^2 \rangle )^n \rangle \propto \langle v^2 \rangle^n$ for $n = 3$, $4$, ..., where $\langle (v_R^2 - \langle v_R^2 \rangle )^n \rangle$ in itself depends on the spatial correlations of $v^2$ among up to $n$ points.\cite{r54} We specify the coefficients of these relations with a thermostatistical formalism [Eq.~(\ref{eq10})].

\begingroup
\squeezetable
\begin{table*}
\caption{\label{t1} 
Experimental conditions and turbulence parameters of grid turbulence (G1, G2, and G3), boundary layer (B1, B2, and B3), and jet (J1, J2, and J3). The velocity derivative was obtained as $\partial _x v = [ 8v(x+\delta x)-8v(x-\delta x)-v(x+2\delta x)+v(x-2\delta x) ] / 12\delta x$ with the sampling interval $\delta x=U/f$.}

\begin{ruledtabular}
\begin{tabular}{lllccccccccc}
\noalign{\smallskip}
Quantity                            &                                         
& Units                             & G1      & G2      & G3      & B1      & B2      & B3      & J1      & J2      & J3   \\ 
\hline
\noalign{\smallskip}
Measurement position                & $x_{\rm wt}$ 
& m                                 & $+1.5$  & $+1.5$  & $+2.0$  & $+12.5$ & $+12.5$ & $+12.5$ & $+15.5$ & $+15.5$ & $+15.5$  \\
Measurement position                & $z_{\rm wt}$ 
& m                                 & $1.00$  & $1.00$  & $1.00$  & $0.35$  & $0.30$  & $0.25$  & $0.40$  & $0.40$  & $0.40$   \\
Sampling frequency                  & $f$ 
& kHz                               & $10$    & $24$    & $54$    & $10$    & $26$    & $60$    & $16$    & $44$    & $90$     \\
Total number of data                &
& $10^8$                            & $1$     & $4$     & $4$     & $1$     & $4$     & $4$     & $1$     & $4$     & $4$      \\     
Kinematic viscosity                 & $\nu$
& cm$^2$\,s$^{-1}$                  & $0.141$ & $0.142$ & $0.143$ & $0.138$ & $0.143$ & $0.143$ & $0.139$ & $0.139$ & $0.139$  \\
Mean velocity                       & $U$
& m\,s$^{-1}$                       & $4.31$  & $8.57$  & $16.9$  & $3.12$  & $5.93$  & $11.3$  & $5.59$  & $11.5$  & $22.9$   \\
Mean energy dissipation             & $\langle \varepsilon \rangle= 15 \nu \langle (\partial _x v)^2 \rangle /2$ 
& m$^2$\,s$^{-3}$                   & $0.141$ & $0.975$ & $4.42$  & $0.244$ & $2.05$  & $12.6$  & $0.379$ & $2.60$  & $15.4$   \\ 
Kolmogorov velocity                 & $u_{\rm K} = (\nu \langle \varepsilon \rangle)^{1/4}$
& m\,s$^{-1}$                       & $0.0375$& $0.0610$& $0.0891$& $0.0428$& $0.0736$& $0.116$ & $0.0479$ & $0.0776$& $0.121$  \\
rms $u$ fluctuation                 & $\langle u^2 \rangle^{1/2}$ 
& m\,s$^{-1}$                       & $0.236$ & $0.475$ & $0.862$ & $0.552$ & $1.18$  & $2.37$  & $0.743$ & $1.56$  & $3.08$   \\
rms $v$ fluctuation                 & $\langle v^2 \rangle^{1/2}$ 
& m\,s$^{-1}$                       & $0.231$ & $0.463$ & $0.837$ & $0.464$ & $0.975$ & $1.96$  & $0.661$ & $1.36$  & $2.71$   \\
Skewness of $u$                     & $\langle u^3 \rangle / \langle u^2 \rangle^{3/2}$
&                                   & $+0.07$ & $+0.05$ & $+0.06$ & $-0.22$ & $-0.12$ & $-0.10$ & $-0.03$ & $-0.04$ & $-0.04$  \\
Skewness of $v$                     & $\langle v^3 \rangle / \langle v^2 \rangle^{3/2}$
&                                   & $-0.01$ & $-0.01$ & $-0.01$ & $+0.01$ & $+0.01$ & $-0.01$ & $-0.01$ & $+0.01$ & $+0.01$  \\
Kurtosis of $u$                     & $\langle u^4 \rangle / \langle u^2 \rangle^2 -3$
&                                   & $-0.00$ & $+0.03$ & $+0.02$ & $-0.31$ & $-0.34$ & $-0.31$ & $-0.40$ & $-0.40$ & $-0.42$   \\
Kurtosis of $v$                     & $\langle v^4 \rangle / \langle v^2 \rangle^2 -3$ 
&                                   & $-0.01$ & $-0.00$ & $+0.00$ & $+0.06$ & $+0.03$ & $+0.05$ & $+0.03$ & $+0.05$ & $+0.06$   \\
Kolmogorov length                   & $\eta = (\nu ^3 / \langle \varepsilon \rangle )^{1/4}$ 
& cm                                & $0.0376$& $0.0233$& $0.0160$& $0.0322$& $0.0194$& $0.0123$& $0.0290$& $0.0179$& $0.0115$ \\
Taylor microscale                   & $\lambda = [2 \langle v^2 \rangle / \langle (\partial _x v)^2 \rangle ]^{1/2}$
& cm                                & $0.896$ & $0.684$ & $0.583$ & $1.35$  & $0.996$ & $0.806$ & $1.55$  & $1.21$  & $0.997$\\
Correlation length of $u$           & $L_u = \int^{\infty}_{0} \langle u(x+r)u(x) \rangle dr / \langle u^2 \rangle$
& cm                                & $16.4$  & $17.0$  & $18.2$  & $49.0$  & $42.4$  & $43.0$  & $130.$  & $128.$  & $128.$   \\
Correlation length of $v$           & $L_v = \int ^{\infty}_{0} \langle v(x+r)v(x) \rangle dr / \langle v^2 \rangle$
& cm                                & $4.13$  & $4.07$  & $4.55$  & $6.94$  & $6.14$  & $5.68$  & $10.3$  & $10.2$  & $10.5$    \\
Correlation length of $u^2$         & $L_{u^2}$ [see Eq.~(\ref{eq1})]
& cm                                & $4.79$  & $4.91$  & $5.40$  & $18.0$  & $15.0$  & $15.0$  & $21.0$  & $21.7$  & $21.9$    \\
Correlation length of $v^2$         & $L_{v^2}$ [see Eq.~(\ref{eq1})]
& cm                                & $2.42$  & $2.49$  & $2.65$  & $8.33$  & $7.22$  & $7.34$  & $12.8$  & $12.9$  & $14.3$    \\
Reynolds number                     & Re$_{\lambda} = \lambda \langle v^2 \rangle^{1/2} / \nu$  
&                                   & $147$   & $223$   & $341$   & $454$   & $679$   & $1103$  & $738$   & $1183$  & $1944$    \\
$U$-$u^2$ correlation at $4L_{u^2}$ & [see Eq.~(\ref{eq9})]
&                                   & $+0.03$ & $+0.01$ & $+0.03$ & $-0.27$ & $-0.18$ & $-0.16$ & $-0.05$ & $-0.07$ & $-0.07$    \\
$U$-$v^2$ correlation at $4L_{v^2}$ & [see Eq.~(\ref{eq9})]
&                                   & $-0.00$ & $-0.01$ & $-0.01$ & $-0.08$ & $-0.04$ & $-0.02$ & $-0.04$ & $-0.04$ & $-0.04$    \\

\end{tabular}
\end{ruledtabular}
\end{table*}
\endgroup

\section{Thermostatistical Formalism} \label{S3}

There is an analogue of Eq.~(\ref{eq3b}) in the statistical mechanics of equilibrium systems with many degrees of freedom. It is a formula for thermal fluctuations of the energy $E$ in a canonical ensemble that has the size $R$ and is in contact with a heat bath at temperature $T$:\cite{k65,c85}
\begin{equation}
\label{eq4}
\langle (E - \langle E \rangle )^2 \rangle = C_R T^2
\quad
\mbox{with}
\quad
C_R = \left( \frac{\partial \langle E \rangle}{\partial T} \right)_R .
\end{equation}
The derivative is taken for constant $R$. We have assumed that $R v_R^2$ at $R \gg L_{v^2}$ is additive. Since $E$ is also additive, Eq.~(\ref{eq4}) is equivalent to Eq.~(\ref{eq3b}) through the correspondences
\begin{subequations}
\label{eq5}
\begin{equation}
\label{eq5a}
T = \frac{\langle v^2 \rangle}{\sqrt{\zeta}}
\quad
\mbox{and}
\quad
E = N \left[ v_R^2 - (1-\sqrt{\zeta}) \langle v^2 \rangle \right] ,
\end{equation}
with
\begin{equation}
\label{eq5b}
N = \frac{R}{4L_{v^2}} \gg 1,
\end{equation}
and hence
\begin{equation}
\label{eq5c}
\langle E \rangle = \zeta NT 
\quad
\mbox{and}
\quad
C_R = \zeta N .
\end{equation}
\end{subequations}
Here $\zeta > 0$ is an arbitrary constant that is to be determined later [Eq.~(\ref{eq8})].

Each segment with length $R$ consists of $N$ subsegments with length $4L_{v^2}$ and mean energy $\sqrt{\zeta} \langle v^2 \rangle$. The adjacent subsegments might be somewhat correlated, but such a correlation has to be negligible at the larger scales. Thus, the segment is a collection of $N$ distinct motions, which are individually attributable to the energy-containing eddies. Once determined, $N = R/4L_{v^2}$ is kept constant even if $R$ varies afterwards [Eq.~(\ref{eq12d})], by assuming that the turbulence expands or contracts in a self-similar manner so that $L_{v^2}$ varies with $R$. The segment is in an equilibrium with the surrounding turbulence that serves as a heat bath at $T = \langle v^2 \rangle / \sqrt{\zeta}$. Although this is not a true temperature, the analogy is close enough to reproduce the observed distribution of $v_R^2$ in Sec.~\ref{S4}.

The energy distribution $P(E)$ in any canonical ensemble is determined by the heat capacity $C_R$,\cite{k65,c85} through a series of basic relations of the statistical mechanics. Since $C_R$ is related to the entropy $\langle S \rangle$ as $C_R = T ( \partial_T \langle S \rangle )_R$, we integrate $C_R = \zeta N$ in Eq. (\ref{eq5c}) to obtain
\begin{subequations}
\label{eq6}
\begin{equation}
\label{eq6a}
\langle S \rangle 
= 
\zeta N \left[ \ln \left( \frac{T}{T_0} \right) +1 \right] ,
\end{equation}
with a constant of integration $\zeta N ( 1 - \ln T_0 )$ that could depend on $R$ via $T_0$. The Helmholtz free energy $\langle F \rangle = \langle E \rangle - T \langle S \rangle$ is 
\begin{equation}
\label{eq6b}
\langle F \rangle = - \zeta N T \ln \left( \frac{T}{T_0} \right) .
\end{equation}
The partition function $Z = \exp ( - \langle F \rangle / T )$ is
\begin{equation}
\label{eq6c}
Z = \left( \frac{T}{T_0} \right)^{\zeta N} .
\end{equation}
\end{subequations}
From the inverse of the Laplace transformation $Z(T) = \int^{\infty}_0 \Omega(E) \exp (-E/T) dE$, we obtain the density of states $\Omega (E) = E^{\zeta N-1} / \Gamma (\zeta N) T_0^{\zeta N}$, where $\Gamma$ is the Gamma function. Lastly, $P(E) = \Omega(E) \exp(-E/T)/Z(T)$ is obtained independently of $T_0$ as
\begin{equation}
\label{eq7}
P(E) = \frac{E^{\zeta N-1}\exp(-E/T)}{\Gamma(\zeta N) T^{\zeta N}} .
\end{equation}
The maximum is at $E = (\zeta N-1)T$. In the limit $N \rightarrow \infty$, the distribution $P(E)$ becomes Gaussian in accordance with the central limit theorem.\cite{f95,k65,c85,f71}

To determine the value of $\zeta$, we assume universality of $P(E)$ at $N \gg 1$. Such universality originates in steps of the energy transfer. They could be related to interactions between the adjacent subsegments. Let us first consider a special case where the $N$ subsegments are not correlated at all but exhibit the same energy distribution that is for the square of a Gaussian random variable. The resultant $P(E)$ at any $N$ is the $\chi^2$ distribution with $N$ degrees of freedom,\cite{f71} which corresponds to Eq.~(\ref{eq7}) with
\begin{equation}
{\textstyle
\label{eq8}
\zeta = \frac{1}{2}.
}
\end{equation}
Then, also at $N \gg 1$ in other general cases, the universality ensures the same value for $\zeta$. Since $\zeta = 1/2$ yields $\langle E \rangle = NT/2$ [Eq.~(\ref{eq5c})], it might be possible to reformulate the large-scale fluctuations in the classical statistical mechanics where $\langle E \rangle = NT/2$ holds as the law of energy equipartition among $N$ degrees of freedom.\cite{k65,ll80,c85}

\begin{figure}[b]
\resizebox{8.3cm}{!}{\includegraphics*[4.5cm,19.9cm][16.5cm,26cm]{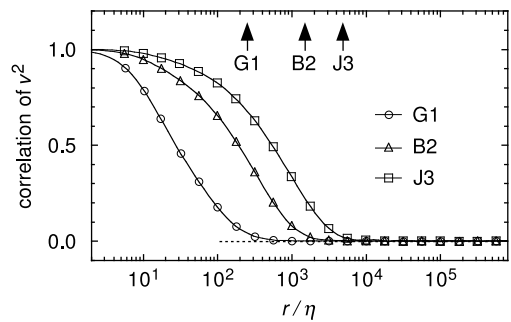}}
\caption{\label{f2} Two-point correlation $\langle [v^2(x+r) - \langle v^2 \rangle][v^2(x) - \langle v^2 \rangle] \rangle$ normalized by its value at $r = 0$ as a function of $r/\eta$ in grid turbulence G1 (circles), boundary layer B2 (triangles), and jet J3 (squares). The arrows indicate $r = 4L_{v^2}$.}
\end{figure} 
\begin{figure}[t]
\resizebox{8.3cm}{!}{\includegraphics*[4.5cm,10.2cm][16.5cm,26cm]{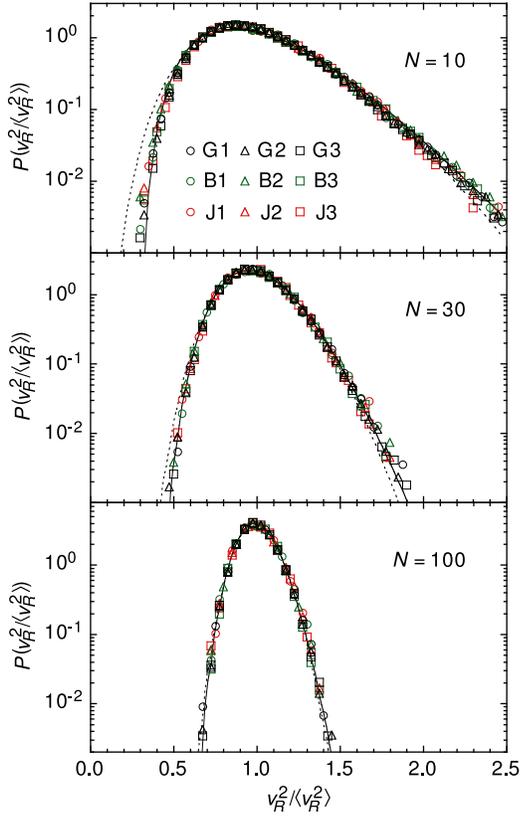}}
\caption{\label{f3} (Color) Probability distribution of $v_R^2 /\langle v_R^2 \rangle$ at $N = R/4L_{v^2} = 10$, $30$, and $100$ in grid turbulence G1, G2, and G3 (black symbols), boundary layer B1, B2, and B3 (green symbols), and jet J1, J2, and J3 (red symbols). The solid and the dotted lines are the theoretical predictions for $\zeta = 1/2$ and $1$.}
\end{figure} 
\begin{figure}[b]
\resizebox{8.3cm}{!}{\includegraphics*[4.5cm,10.4cm][16.5cm,26cm]{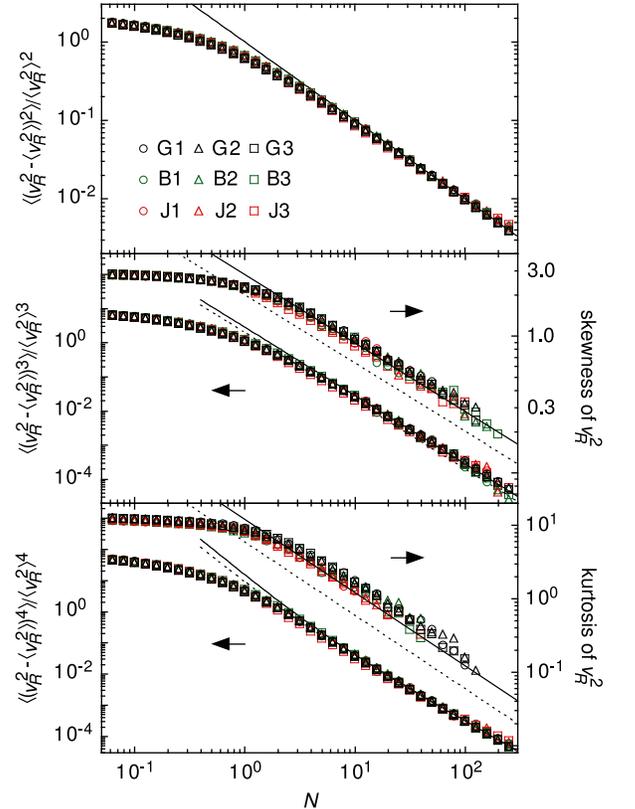}}
\caption{\label{f4} (Color) Moments $\langle (v_R^2-\langle v_R^2 \rangle )^n \rangle$ normalized by $\langle v_R^2 \rangle^n$ for $n = 2$, $3$, and $4$ as a function of $N = R/4L_{v^2}$. We also show the skewness $\langle (v_R^2- \langle v_R^2 \rangle)^3 \rangle / \langle (v_R^2- \langle v_R^2 \rangle)^2 \rangle^{3/2}$ and the kurtosis $\langle (v_R^2- \langle v_R^2 \rangle)^4 \rangle / \langle (v_R^2- \langle v_R^2 \rangle)^2 \rangle^{2}-3$, as long as the value is reliable. The symbols are the same as in Fig.~\ref{f3}. In the top panel, the solid line matches the dotted line.}
\end{figure} 

\section{Confirmation by Experiments} \label{S4}

The theoretical distribution of $E$ [Eq.~(\ref{eq7})] leads to the distribution of $v_R^2$, which is confirmed with experimental data of grid turbulence (G1, G2, and G3), boundary layer (B1, B2, and B3), and jet (J1, J2, and J3). While B1, B3, and J2 were used in our past work,\cite{mht09} the others are used here for the first time. Their conditions and turbulence parameters are summarized in Table~\ref{t1}.

The experiments were conducted under stationary conditions in a wind tunnel, which had a test section of the size of $18 \times 3 \times 2$\,m$^3$. At a position where the turbulence was fully developed, we measured the streamwise velocity $U+u(t_{\rm wt})$ and the spanwise velocity $v(t_{\rm wt})$. Here $U$ is the average while $u(t_{\rm wt})$ and $v(t_{\rm wt})$ are temporal fluctuations. They were converted into spatial fluctuations of the longitudinal velocity $u(x)$ and of the lateral velocity $v(x)$ by using Taylor's hypothesis, $x = -U t_{\rm wt}$. For further details of the experiments, see Appendix.

Figure~\ref{f2} shows examples of the two-point correlation of the local energy of the lateral velocity $v^2$, which is used to obtain the subsegment length $4L_{v^2}$ [Eq.~(\ref{eq1})]. The correlation appears to be almost negligible above the scale of $4 L_{v^2}$ (arrows) as assumed in our formalism. For a range of $N = R/4L_{v^2}$, we calculate the coarse-grained energy $v_R^2$ in each segment with length $R$ [Eq.~(\ref{eq2})]. That of the longitudinal velocity $u$ is to be studied in Sec.~\ref{S6}.

Having the length $4 L_{v^2} \simeq 0.1$--$0.6$\,m (Table~\ref{t1}), the subsegments of $v(x)$ are local enough to represent local regions of stationary turbulence that did exist in the wind tunnel. The adjacent regions were interacting. We have connected these subsegments to make up segments with any length $R$. Although the turbulence in the wind tunnel was limited up to the length of its test section, no inconsistency arises because each of the segments is used not as a single motion with length $R$ but as a collection of $N$ distinct motions with length $4 L_{v^2}$. In fact, the statistical mechanics allows us to make up a canonical system by collecting subsystems that might be even isolated from one another,\cite{c85} i.e., our subsegments, if they are at the same temperature, i.e., $T = \langle v^2 \rangle / \sqrt{\zeta}$ [Eq.~(\ref{eq5a})]. The segment is also homogeneous in the sense that its subsegments all obey the same statistical law. Since the total number of the subsegments is as large as $10^5$--$10^6$ in the individual data records (Table~\ref{t1}), the resulting statistics are expected to be reliable.

The mean streamwise velocity $U_R$ in each segment with length $R$ is not constant at the value of the mean velocity for the entire data, $U = \langle U_R \rangle$.\cite{kkp98} To confirm that such fluctuations of $U_R$ do not affect our study of $v_R^2$, we calculate their correlation
\begin{equation}
\label{eq9}
\frac{\langle (U_R -\langle U_R \rangle)                         (v^2_R -\langle v^2_R \rangle )   \rangle}
     {\langle (U_R -\langle U_R \rangle)^2 \rangle^{1/2} \langle (v^2_R -\langle v^2_R \rangle )^2 \rangle^{1/2}} ,
\end{equation}
which is surely negligible at the scale of the subsegment length, $R = 4 L_{v^2}$ (Table~\ref{t1}). At around this scale, $\langle (U_R -\langle U_R \rangle)^2 \rangle$ is $70$\% of $\langle u^2 \rangle$.

Figure~\ref{f3} shows the probability distribution of $v_R^2/\langle v_R^2 \rangle$ at $N = R/4L_{v^2} = 10$, $30$, and $100$. The solid and the dotted lines are the theoretical predictions of Eq.~(\ref{eq7}) via Eq. (\ref{eq5a}) for $\zeta = 1/2$ and $1$, which depend on $N$ alone. With an increase in $N$, the distribution becomes narrower, but it remains wide enough to imply the significance of the fluctuations.\cite{po97,c03,m06,mht09} The experiments agree with one another and with the theory for $\zeta = 1/2$ [Eq.~(\ref{eq8})].

Figure~\ref{f4} shows $\langle (v_R^2-\langle v_R^2 \rangle )^n \rangle$ normalized by $\langle v_R^2 \rangle^n$ for $n = 2$, $3$, and $4$ as a function of $N = R/4L_{v^2}$. Also shown are the skewness $\langle (v_R^2- \langle v_R^2 \rangle)^3 \rangle / \langle (v_R^2- \langle v_R^2 \rangle)^2 \rangle^{3/2}$ and the kurtosis $\langle (v_R^2- \langle v_R^2 \rangle)^4 \rangle / \langle (v_R^2- \langle v_R^2 \rangle)^2 \rangle^{2}-3$. Theoretically, Eq.~(\ref{eq7}) yields
\begin{subequations}
\label{eq10}
\begin{eqnarray}
&& \label{eq10a} \langle (E- \langle E \rangle)^2 \rangle = \zeta NT^2 ,  \\
&& \label{eq10b} \langle (E- \langle E \rangle)^3 \rangle = 2\zeta N T^3 ,\\
&& \label{eq10c} \langle (E- \langle E \rangle)^4 \rangle = (3 \zeta ^2 N^2+6 \zeta N)T^4 ,
\end{eqnarray}
\end{subequations}
with $\langle E \rangle = \zeta NT$ [Eq.~(\ref{eq5c})]. They are used with Eq.~(\ref{eq5a}) to obtain the relations of $v_R^2$ for $\zeta = 1/2$ and $1$ (solid and dotted lines), which depend on $N$ alone. Again at $N \gtrsim 10^1$, the experiments agree with one another and with the theory for $\zeta = 1/2$ [Eq.~(\ref{eq8})].

Thus, we have reproduced the experiments of the stationary homogeneous turbulence. Negligible, if any, are undesired variations in conditions of the wind tunnel and of the measurement devices. They do not explain the observed agreement among the experiments, for which $v_R^2$ and $R$ have been normalized by $\langle v^2 \rangle$ and $4 L_{v^2}$, i.e., characteristics of the turbulence. We have thereby confirmed our thermostatistical formalism as well as its assumptions that $R v_R^2$ is additive and $v_R^2$ has a universal distribution at $R \gg L_{v^2}$.

\section{Complete Formalism and its Implication} \label{S5}

To complete the thermostatistical formalism, we determine $T/T_0$ in Eq.~(\ref{eq6}). This quantity is dimensionless and characterizes the turbulence in the subsegments. Judging from $T = \langle v^2 \rangle / \sqrt{\zeta}$ [Eq.~(\ref{eq5a})], it is natural to equate $T/T_0$ with the square of the Reynolds number $\mbox{Re}_{4L}^2$ for those subsegments with length $4L_{v^2}$: 
\begin{equation}
\label{eq11}
\mbox{Re}_{4L}^2 = \frac{16 L_{v^2}^2 \langle v^2 \rangle}{\nu^2} = \frac{\sqrt{\zeta} R^2 T}{N^2 \nu^2}.
\end{equation}
Here $\nu$ is the kinematic viscosity. For $\zeta = 1/2$ [Eq.~(\ref{eq8})], the partition function $Z$ in Eq.~(\ref{eq6c}) becomes
\begin{subequations}
\label{eq12}
\begin{equation}
\label{eq12a}
Z = \mbox{Re}_{4L}^N = \left( \frac{R^2 T}{\sqrt{2} N^2 \nu^2} \right)^{N/2} .
\end{equation}
The Helmholtz free energy $\langle F \rangle$ in Eq.~(\ref{eq6b}) becomes
\begin{equation}
\label{eq12b}
\langle F \rangle 
=
-\frac{NT}{2} \ln \left( \frac{R^2T}{\sqrt{2} N^2\nu^2} \right) .
\end{equation}
The entropy $\langle S \rangle$ in Eq.~(\ref{eq6a}) becomes 
\begin{equation}
\label{eq12c}
\langle S \rangle 
= 
- \left( \frac{\partial \langle F \rangle}{\partial T} \right)_R
=
\frac{N}{2} \left[ \ln \left( \frac{R^2 T}{\sqrt{2} N^2 \nu^2} \right) +1 \right].
\end{equation}
Being equivalent to $\ln ( e \mbox{Re}_{4L}^2 )^{N/2}$, the entropy $\langle S \rangle$ is large if the Reynolds number $\mbox{Re}_{4L}$ is high. Lastly, the resistance force $\langle f \rangle$ is obtained as
\begin{equation}
\label{eq12d}
\langle f \rangle
=
- \left( \frac{\partial \langle F \rangle}{\partial R} \right)_T
=
\frac{NT}{R} .
\end{equation}
\end{subequations}
This is analogous to a force originating in the Reynolds stress, $\partial_{x_j} \langle v_i v_j \rangle$ for the velocity $v_i$ in a direction $x_i$. The reason is $NT/R \propto \langle v^2 \rangle / L_{v^2}$, where $L_{v^2}$ serves as the scale for a significant variation of $v^2$.

Our formalism of Eq.~(\ref{eq12}) agrees with the thermodynamics. While $T$ and $\langle f \rangle$ are intensive, $\langle S \rangle$, $R$, and $\langle F \rangle$ are extensive, $\propto N$. Through the Legendre transformation, $\langle F \rangle$ yields other thermodynamic potentials,\cite{ll80,c85} e.g., the Gibbs free energy $\langle G \rangle = \langle F \rangle + \langle f \rangle R$ as a function of $T$ and $\langle f \rangle$:
\begin{equation}
\label{eq13}
\langle G \rangle
=
\frac{NT}{2} \left[ 2- \ln \left( \frac{T^3}{\sqrt{2} \nu^2 \langle f \rangle^2} \right) \right] ,
\end{equation}
with $\langle S \rangle = - ( \partial_T \langle G \rangle )_{\langle f \rangle}$. These potentials are totally differentiable and hence reproduce the Maxwell relations.\cite{ll80,c85} For example, from $[ \partial_R ( \partial_T \langle F \rangle )_R ]_T = [ \partial_T ( \partial_R \langle F \rangle )_T ]_R$ in Eq.~(\ref{eq12}), we have
\begin{subequations}
\begin{equation}
\label{eq14a}
\left( \frac{\partial \langle S \rangle}{\partial R} \right)_T 
=
\left( \frac{\partial \langle f \rangle}{\partial T} \right)_R .
\end{equation}
To describe an equilibrium state, the potentials also reproduce the so-called thermodynamic inequalities.\cite{ll80,c85} For example, between $C_R = T ( \partial_T \langle S \rangle )_R$ from $\langle F \rangle$ in Eq.~(\ref{eq12}) and $C_{\langle f \rangle} = T ( \partial_T \langle S \rangle )_{\langle f \rangle}$ from $\langle G \rangle$ in Eq.~(\ref{eq13}), we have a relation analogous to a well-known inequality\cite{ll80,c85} between heat capacities at constant pressure and at constant volume:
\begin{equation}
\label{eq14b}
C_{\langle f \rangle} = \frac{3N}{2} > C_R = \frac{N}{2} > 0. 
\end{equation}
\end{subequations}
This agreement with the thermodynamics implies that our formalism is surely thermostatistical. The conclusion remains the same even if $\zeta \ne 1/2$, for which we only have to consider $Z = \mbox{Re}_{4L}^{2 \zeta N}$ and so on.

\section{Concluding Discussion} \label{S6}

For stationary and homogeneous turbulence, we have studied large-scale fluctuations of the coarse-grained energy of the lateral velocity $v_R^2$ [Eq.~(\ref{eq2})]. They have been described in a thermostatistical formalism, which has the same mathematical structure as used for the statistical mechanics of equilibrium systems with many degrees of freedom. By using an analogy between the fluctuations of $v_R^2$ [Eq.~(\ref{eq3b})] and the thermal fluctuations of the energy $E$ [Eq.~(\ref{eq4})], we have obtained a correspondence between $v_R^2$ and $E$ [Eq.~(\ref{eq5})]. The resultant formalism reproduces the distribution of $v_R^2$ observed at $N = R/4L_{v^2} \gtrsim 10^1$ in Figs.~\ref{f3} and \ref{f4} [Eqs.~(\ref{eq7}) and (\ref{eq8})]. Therefore, through the large-scale fluctuations, turbulence is related to the statistical mechanics.

The thermostatistical formalism of Onsager for a class of two-dimensional turbulence is well known.\cite{f95,o49,es06} We have demonstrated that such a formalism also exists at large scales of the usual three-dimensional turbulence, although we have used a canonical ensemble while Onsager used a microcanonical ensemble\cite{k65,ll80,c85} at constant $E$.

To construct the formalism, we have assumed that $R v_R^2$ is additive at $R \gg L_{v^2}$ [Eqs.~(\ref{eq3b}) and (\ref{eq5a})], by assuming that any $n$-point spatial correlation of $v^2$ is negligible at $r \gg L_{v^2}$. We have also assumed that the distribution of $v_R^2$ is universal at $R \gg L_{v^2}$ [Eq.~(\ref{eq8})], by assuming that the dependence on the flow configuration is lost after many random steps of the energy transfer to the large scales. These assumptions have been confirmed along with the formalism. Especially in Figs.~\ref{f3} and \ref{f4}, we have observed the universal distribution of $v_R^2$.

However, the additivity and the universality might not be exact. The spatial correlations of $v^2$ might not be exactly negligible in flows such as those known to have organized motions far above the scale of the energy supply $L$,\cite{a07,mhnmc09} where the additivity might be lost in some manner specific to the flow configuration. Since the effective degrees of freedom of the segment might be less than $N$, the effective value of $\zeta$ might be less than $1/2$ if we recall the discussion leading to Eq.~(\ref{eq8}). Thus, our formalism might not be exact. Nevertheless, it is at least a good approximation because we have observed no disagreement with the experiments. Similar discussions exist about possible effects of the flow configuration on otherwise universal motions at small scales.\cite{f95,m06,ll59,k74,ks00}

\begin{figure}[t]
\resizebox{8.3cm}{!}{\includegraphics*[4.5cm,10.4cm][16.5cm,26cm]{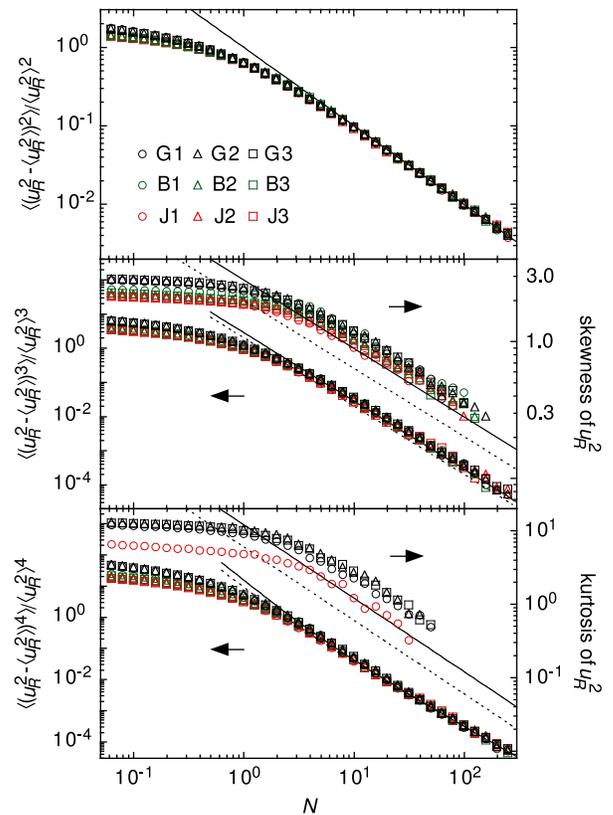}}
\caption{\label{f5} (Color) Same as in Fig.~\ref{f4} but for $u_R^2$ as a function of $N = R/4L_{u^2}$. For the kurtosis, we only show the values of G1, G2, G3, and J1. These are the most reliable experiments where the correlation of $U_R$ with $u_R^2$ at $R = 4L_{u^2}$ is negligible (Table~\ref{t1}). Between the values of G1--G3 and J1, scattered are the values of the other experiments.}
\end{figure}

The additivity and the universality have not been confirmed for the coarse-grained energy of the longitudinal velocity $u_R^2$. Figure~\ref{f5} shows that our formalism does not reproduce the experiments of some of the flows. Their skewness and kurtosis, among others, are greater than those for $\zeta = 1/2$ (solid lines), implying that their effective values of $\zeta$ are significantly less than $1/2$. These are significant versions of the above mentioned feature. Presumably, along the one-dimensional cut of an incompressible fluid, longitudinal distortions propagate to larger distances than lateral distortions. The spatial correlations of the local energy $u^2$ do not always decay fast enough to become negligible at the scales studied here. Still at the larger scales, there remains a possibility to reproduce experiments of all the flows.

Our formalism does not hold at $R \lesssim 4L_{v^2}$, but an approximation is available. For example, if the distribution of $v$ is Gaussian, we have $\langle ( v_R^2 - \langle v_R^2 \rangle )^2 \rangle \rightarrow 2 \langle v^2 \rangle^2$ in the limit $R \rightarrow 0$. Then, Eq.~(\ref{eq3a}) is approximated at any $R$ as $\langle ( v_R^2 - \langle v_R^2 \rangle )^2 \rangle = 4L_{v^2} \langle v^2 \rangle^2 / (R + 2L_{v^2})$. By comparing this with Eq.~(\ref{eq4}), we obtain correspondences as in Eq.~(\ref{eq5}) and a formalism as in Eqs.~(\ref{eq6})--(\ref{eq8}) and (\ref{eq12}).

There is also an application of our formulation to large-scale fluctuations other than those of $v_R^2$, if they are in an equilibrium as well as have additivity and are thereby analogous to thermal fluctuations in the statistical mechanics. We only have to write the mean square in the form of Eq.~(\ref{eq3b}) and compare it with Eq.~(\ref{eq4}). The formulation could be constrained by some additional feature, e.g., universality in Eq.~(\ref{eq8}). This is the case not only for one-dimensional data as studied here but also for any data of the higher dimension. Examples are expected to be found in a variety of fluctuations, far beyond those of turbulence.

\begin{acknowledgments}
This work was supported in part by KAKENHI Grant No. 22540402.
\end{acknowledgments}

\appendix*
\section{DETAILS OF EXPERIMENTS} \label{app}

The experiments were conducted in a wind tunnel of the Meteorological Research Institute. We adopt coordinates $x_{\rm wt}$, $y_{\rm wt}$, and $z_{\rm wt}$ in the streamwise, spanwise, and floor-normal directions. The origin $x_{\rm wt} = y_{\rm wt} = z_{\rm wt} = 0$\,m is on the floor center at the upstream end of the test section of the wind tunnel. Its size was $\delta x_{\rm wt} = 18$\,m, $\delta y_{\rm wt} = 3$\,m, and $\delta z_{\rm wt} = 2$\,m. The cross section $\delta y_{\rm wt} \times \delta z_{\rm wt}$ was the same upstream to $x_{\rm wt} = -4$\,m.

The wind tunnel had an air conditioner. If needed, we used this conditioner to constrain the variation of the air temperature. The resultant variation was $\pm 1\,^{\circ}$C at most in each experiment, where the kinematic viscosity $\nu$ is assumed to have been constant.

To simultaneously measure $U+u$ and $v$, we used a hot-wire anemometer. The anemometer was composed of a constant temperature system and a crossed-wire probe. The wires were made of platinum-plated tungsten, $5$\,$\mu$m in diameter, $1.25$\,mm in sensing length, $1$\,mm in separation, oriented at $\pm 45^{\circ}$ to the streamwise direction, and $280$\,$^{\circ}$C in temperature.

For the grid turbulence, we placed a grid at $x_{\rm wt} = -2$ m across the flow passage to the test section of the wind tunnel. The grid had two layers of uniformly spaced rods, with axes in the two layers at right angles. The cross section of the rod was $0.04 \times 0.04$\,m$^2$. The spacing of the rod axes was $0.20$\,m. We set the incoming flow velocity to be $4$ (G1), $8$ (G2), or $16$\,m\,s$^{-1}$ (G3). The measurement position was on the tunnel axis, $y_{\rm wt} = 0$\,m and $z_{\rm wt} = 1.00$\,m (Table~\ref{t1}).

For the boundary layer, roughness blocks were placed over the entire floor of the test section. The block size was $\delta x_{\rm wt} = 0.06$\,m, $\delta y_{\rm wt} = 0.21$\,m, and $\delta z_{\rm wt} = 0.11$\,m. The spacing of the block centers was $\delta x_{\rm wt} = \delta y_{\rm wt} = 0.50$\,m. We set the incoming flow velocity to be $4$ (B1), $8$ (B2), or $16$\,m\,s$^{-1}$ (B3). The measurement position was in the log-law sublayer at $x_{\rm wt} = +12.5$\,m and $ y_{\rm wt} = 0$\,m (Table~\ref{t1}), where the boundary layer had the displacement thickness of $0.2$\,m and the $99$\% velocity thickness of $0.8$\,m.

For the jet, we placed a contraction nozzle. Its exit was at $x_{\rm wt} = -2$\,m and was rectangular with the size of $\delta y_{\rm wt} = 2.1$\,m and $\delta z_{\rm wt} = 1.4$\,m. The center was on the tunnel axis. We set the flow velocity at the nozzle exit to be $8$ (J1), $16$ (J2), or $33$\,m\,s$^{-1}$ (J3). The measurement position was at $x_{\rm wt} = +15.5$\,m, $y_{\rm wt} = 0$\,m, and $z_{\rm wt} = 0.40$\,m.

These measurement positions were determined so that the skewness $\langle v^3 \rangle / \langle v^2 \rangle^{3/2}$ and the kurtosis $\langle v^4 \rangle / \langle v^2 \rangle^2 -3$ were close to the Gaussian value of 0 (Table~\ref{t1}). It ensures that the turbulence was fully developed and various eddies filled the space randomly and independently.\cite{m02,m03} Not always close to the Gaussian value were $\langle u^3 \rangle / \langle u^2 \rangle^{3/2}$ and $\langle u^4 \rangle / \langle u^2 \rangle^2-3$ (Table~\ref{t1}). They are sensitive to specific features of the energy-containing eddies that depend on the grid, the roughness, or the nozzle.

The signal of the anemometer was linearized, low-pass filtered, and then digitally sampled. We set the sampling frequency as high as possible (Table~\ref{t1}), on the condition that high-frequency noise was not significant in the power spectrum. The filter cutoff was at one-half of the sampling frequency. We obtained a long record of $1 \times 10^8$ or $4 \times 10^8$ data in each of the experiments (Table~\ref{t1}).

The sampled signal is proportional to the flow velocity, through the calibration coefficient that depends on the condition of the anemometer and thereby varied slowly in time. For individual segments of each data record, the length of which is fixed for the record and ranges from $4 \times 10^6$ to $2 \times 10^7$ data, we determined the values of the coefficient so as to have the same $U$ value. The coefficient within any of these segments is estimated to have varied by $\pm 1$\% at most.

\end{document}